\documentclass[12pt, titlepage, reqno]{article}
 \usepackage[super]{natbib}
 \usepackage{amssymb, latexsym}
 \usepackage{amsmath} 
 \usepackage{amsthm}
 \usepackage{bm}
 \usepackage[mathscr]{eucal}
 \usepackage{enumerate}
 \usepackage{subfigure}
\usepackage{dcolumn}
\usepackage{amsmath}
\usepackage{bm}

\usepackage{graphicx}
\usepackage{epstopdf}
\usepackage{tabularx}%

 \renewcommand{\section}[1]{\medskip \addtocounter{section}{1}\raggedright 
     \textbf{\Roman{section}. \ #1}\medskip \setcounter{subsection}{0}
    \setlength{\parindent}{5ex}
 }
 \renewcommand{\subsection}[1]{\medskip \addtocounter{subsection}{1}\raggedright
    \textbf{\Alph{subsection}. \ #1} \medskip \setcounter{subsubsection}{0}\setlength{\parindent}{5ex}
}

 \theoremstyle{plain}
\begin{document}

 \begin{titlepage}

 \begin{center}

 \textbf{Speech Enhancement}\\
 
 \textbf{via Two-Stage Dual Tree Complex Wavelet Packet Transform }\\
 
 \textbf{with a Speech Presence Probability Estimator} \\

 \vspace{10ex}

Pengfei Sun and Jun Qin 

 \end{center}

 \end{titlepage}

 \begin{abstract}

In this paper, a two-stage dual tree complex wavelet packet transform (DTCWPT) based speech enhancement algorithm has been proposed, in which a speech presence probability (SPP) estimator and a generalized minimum mean squared error (MMSE) estimator are developed. To overcome the drawback of signal distortions caused by down sampling of WPT, a two-stage analytic decomposition concatenating undecimated WPT (UWPT) and decimated WPT is employed. An SPP estimator in the DTCWPT domain is derived based on a generalized Gamma distribution of speech, and Gaussian noise assumption. The validation results show that the proposed algorithm can obtain enhanced perceptual evaluation of speech quality (PESQ) , and segmental signal-to-noise ratio (SegSNR) at low SNR nonstationary noise, compared with other four state-of-the-art speech enhancement algorithms, including optimally modified LSA (OM-LSA), soft masking using a posteriori SNR uncertainty (SMPO), a posteriori SPP based MMSE estimation (MMSE-SPP), and adaptive Bayesian wavelet thresholding (BWT).

 \end{abstract}

 \addtocounter{page}{2}


 \section{INTRODUCTION}
 
 \setlength{\parindent}{5ex}
 
Single-channel speech enhancement technologies are able to reduce or suppress background noise in order to improve the quality and intelligibility of speech \citep{loizou2013speech}. They have been widely used in many applications, including hearing aids, robust speech recognition, mobile speech communication, etc.\citep{kim2011gain}. 

To achieve robust and adaptive noise cancellation, various speech enhancement algorithms have been developed over the past two decades. Short time Fourier transform (STFT) based minimum mean squared error (MMSE) spectral amplitude estimation\citep{gerkmann2012unbiased} and discrete wavelet transform (DWT) based subband shrinkage \cite{hu2004speech,bahoura2001wavelet} are two major speech enhancement technologies. For STFT-based MMSE spectral amplitude estimation, numerous studies have been undertaken to improve the statistical model of speech \citep{erkelens2007minimum}, magnitude estimators\citep{cohen2004speech,gerkmann2014mmse}, and speech presence probability (SPP) estimation \citep{cohen2001speech}. In DWT-shrinkage, different threshold strategies \citep{bahoura2001wavelet, weickert2009analytic, lun2012wavelet} are developed to achieve adaptive masking. As an energy-concentrated decomposition, DWT can facilitate the separation of speech and noise components in each subband. Although both MMSE-based and wavelet-shrinkage-based algorithms show improvements of speech enhancement technologies, their performances significantly decrease under low SNR environments \citep{loizou2013speech}. 

To take advantage of MMSE estimation and DWT techniques, a natural way \citep{tacsmaz2008speech} is to apply statistical estimators to DWT coefficients, in which the statistical estimator provides an overall evaluation of the speech and noise distributions. Wavelets, as taper windows to concentrate the energy of speech components and average the periodogram\citep{hu2004speech, schwerin2014using}, have been frequently used to enhance the low SNR speech estimation. However, there are several limitations on DWT decomposition, such as shift-variance, aliasing, and oscillation \citep{weickert2009analytic}. 

Recently, dual tree complex wavelet packet transform (DTCWPT) with the characteristics of shift-invariance and non-oscillation, has been employed in many signal processing frameworks \citep{ weickert2009analytic, tasmaz2015speech}. Comparatively, undecimated DWT (UDWT) \citep{lang1996noise, cohen2001enhancement} is a typical method to overcome the aliasing problem. Therefore, concatenating decimated and undecimated DTCWPT can be a potential solution to these issues (i.e., shift-variance, aliasing, and oscillation) without introducing more redundancy. Furthermore, MMSE estimators are often derived with the assumption that speech is actually present, and previous work shows that incorporating SPP estimators could significantly improve the performance of clean-speech estimators \citep{gerkmann2008improved, cohen2003noise}. To achieve accurate estimation of SPP, different probabilistic latent component based models have been developed\citep{lun2012wavelet, cohen2002optimal}, and most of them are derived in STFT domain. In addition, wavelets as a multitaper function, providing spectrum smoothing effect, has also been used for SPP estimation \citep{lun2012wavelet, sun2016wavelet}. 
 
In this study, we propose a new speech enhancement algorithm by applying an SPP-incorporated MMSE estimator to two-stage DTCWPT coefficients. Employing DTCWPT to decompose the noisy speech is motivated by two considerations: 1) shift-invariant and window tapered spectrums help estimate the speech presence probability accurately, and 2) energy-concentrated representation of clean speech could improve the performance of the MMSE estimator \citep{weickert2009analytic, bayram2008on}. The underlying rationality of the developed two-stage decimated and undecimated structure is that it aids in the elimination of aliasing. An illustrative speech model based on generalized Gamma distribution \citep{erkelens2007minimum} is introduced to develop a more speech-restrictive MMSE estimator, and the parameters of this speech probability model are optimized in terms of the DTCWPT coefficients. Specifically, the incorporated SPP estimator is obtained in the optimized speech model framework. The performance of the proposed algorithm has been evaluated by comparing it with other state-of-the-art speech enhancement algorithms, including optimally modified LSA (OM-LSA)\citep{cohen2003noise}, soft masking using posterior SNR uncertainty (SMPO)\citep{lu2011estimators}, posterior SPP based MMSE estimation (MMSE-SPP)\citep{hendriks2013dft}, and adaptive Bayesian wavelet thresholding (BWT) \citep{chang2000adaptive}.
 \medskip

\section{MATERIALS AND METHODS}

\medskip

\subsection{\label{sec:level2}Two-stage dual tree complex wavelet packets analysis}

\medskip
As shown in Fig.1, the proposed two-stage DTCWPT consists of two branches of wavelet transform, in which the real part of the wavelet coefficients $\mathcal{X}^{\Re{e}}$ and the imaginary part $\mathcal{X}^{\Im{m}}$ are calculated in parallel. We denote the wavelet associated with the first wavelet perfect reconstruction filter bank (FB) as $\psi(t)$ and the wavelet associated with the second FB as $\psi^{'}(t)$ \citep{bayram2008on}. The wavelet $\psi(t)$ is given as 
\begin{equation}
\psi(t) = \sqrt{2}\sum_{n} h_{1}(n)\phi(2t-n)
\end{equation}
where 
\begin{equation}
\phi(t) = \sqrt{2}\sum_{n} h_{0}(n)\phi(2t-n)
\end{equation}
The second wavelet $\psi^{'}(t)$ is defined similarly in terms of $\{h_{0}^{'}(n),h_{1}^{'}(n)\}$. Specifically, $\{h_{0}(n),h_{1}(n)\}$ is assumed to form an FIR conjugate quadrature filter pair, as does $\{h_{0}^{'}(n),h_{1}^{'}(n)\}$. The outputs of the two trees representing the $\Re{e}$ part and $\Im{m}$ part of the DTCWPT coefficients can be defined as
\begin{equation}
\mathcal{X}_{l}(t) = \sum_{n= -\infty}^{\infty}H_{C}^{l}(t-n) x(n)
\nonumber
\end{equation}
\begin{equation}
  = \sum_{n=-\infty}^{\infty} \mid H_{C}^{l}(t-n) \mid x(n) e^{-j\phi_{C}^{l}(n)}       
\end{equation}
where $ H_{C}^{l} =H^{l}+jH^{(l)'}$, and $H^{l}$ and $H^{(l)'}$ refer to the multiple filter combinations at tree $\Re{e}$ and tree $\Im{m}$, respectively. $l$ refers to the level of wavelet decomposition, and $t$ is the time index. In (3), the dual tree wavelet $H_{C}^{l}$ can be regarded as a Fourier transform succeeded with a taper window $\mid H_{C}^{l} \mid$. Similar to a Hanning window, the bandpass filter $\mid H_{C}^{l}(n) \mid$ excludes higher and lower harmonics. When $l\geq 6$ and with an 8 kHz sampling rate in this study, the bandwidth of each subband is less than 62.5 Hz, which is lower than most of the fundamental frequencies of human tones. In other words, for each wavelet band, only a single harmonic component of speech contributes to the amplitude. Practically, in our study we select $l=7$, in which 4 levels are for decimated DTCWPT, and 3 levels are for undecimated DTCWPT. Therefore, \(2^{7}\) subbands provide enough frequency resolution, and three-level undecimated DTCWPT prevents aliasing caused by down sampling. The remaining four-level DTCWPT helps to avoid redundancy.

In the DTCWPT domain, the noisy speech magnitude is an additive superposition of speech components and noise components,
\begin{equation}
X_{l}(t) = A_{l}(t) + W_{l}(t)
\end{equation}

For notational convenience, the time-frame index \(t\) and subband index \(l\) will be left out. The advantages of the proposed 2-stage DTCWPT have been demonstrated in Fig.2. In the time period [0.8 1.2], DWT coefficients are highly oscillated compared with DTCWPT and the proposed 2-stage DTCWPT coefficients. As a result, for SPP estimation, it would be difficult to detect the presence of speech in the interval 0.8 to 1.2 s using DWT coefficients. Comparatively, the proposed 2-stage DTCWPT provides much stronger and more stable peaks than DTCWPT by solving the aliasing problem. The redundancy caused by undecimated transformation leads to local high frequency components as shown in Fig.2(c), which basically oscillate in a narrow magnitude range and show little impact on speech detection. This phenomenon further proves the rationality of the 2-stage concatenated structure: on one hand, it introduces slight redundancy to obtain strong speech peaks. On the other hand, by decimated decomposition, the local oscillation will be smoothed to avoid high level artificial transitions.   

\begin{figure}[!hbt]
\vspace{-2mm}
\centerline{\includegraphics[scale=0.90]{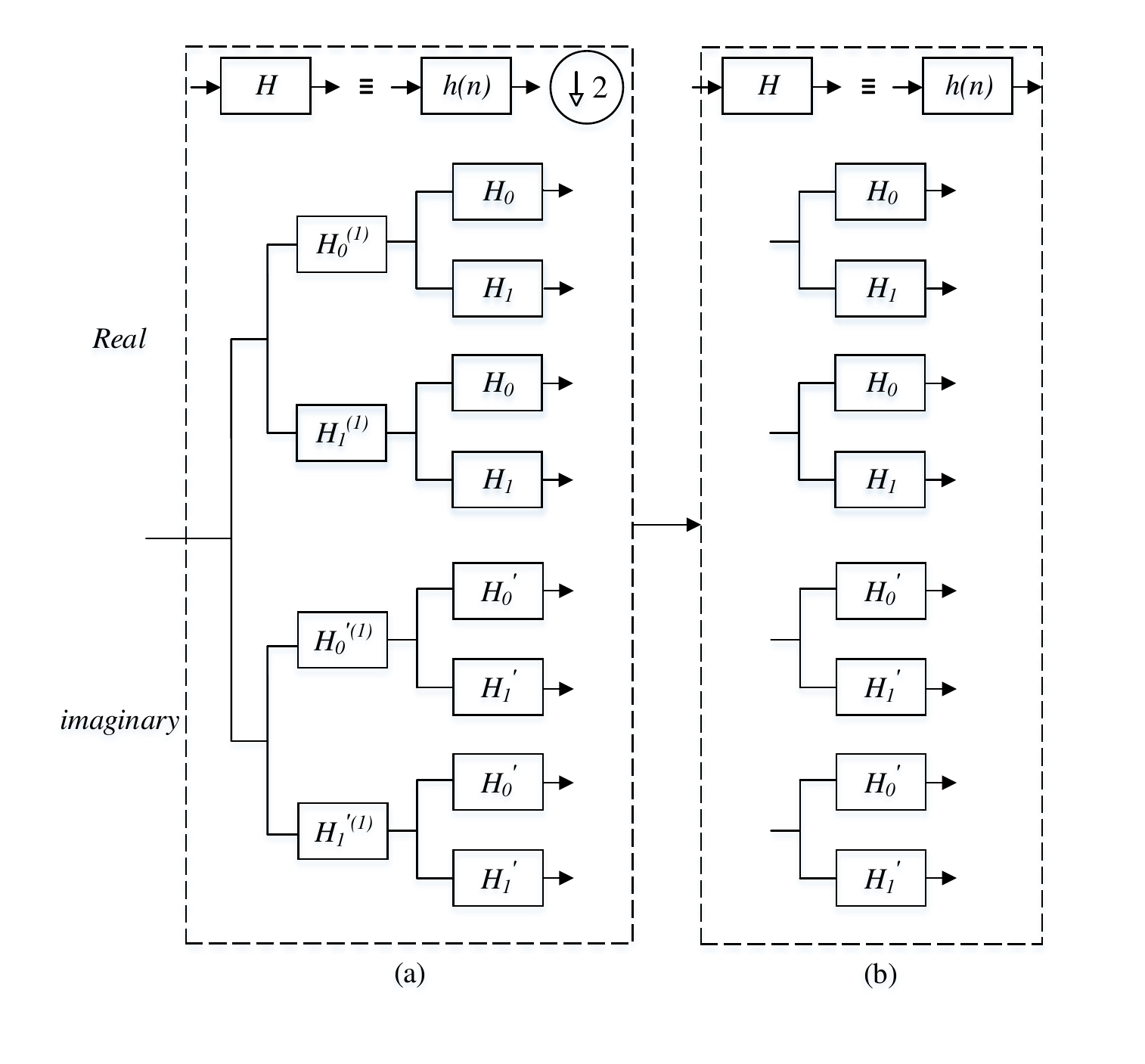}}
\caption{The concatenation of the (a) decimated and (b) undecimated DTCWPT. For each level decomposition, the down sampling reduces the data size into half length in decimated DTCWPT, whereas no down sampling for undecimated DTCWPT.}
\end{figure}
\begin{figure}[!hbt]
\vspace{-2mm}
\centerline{\includegraphics[scale=0.90]{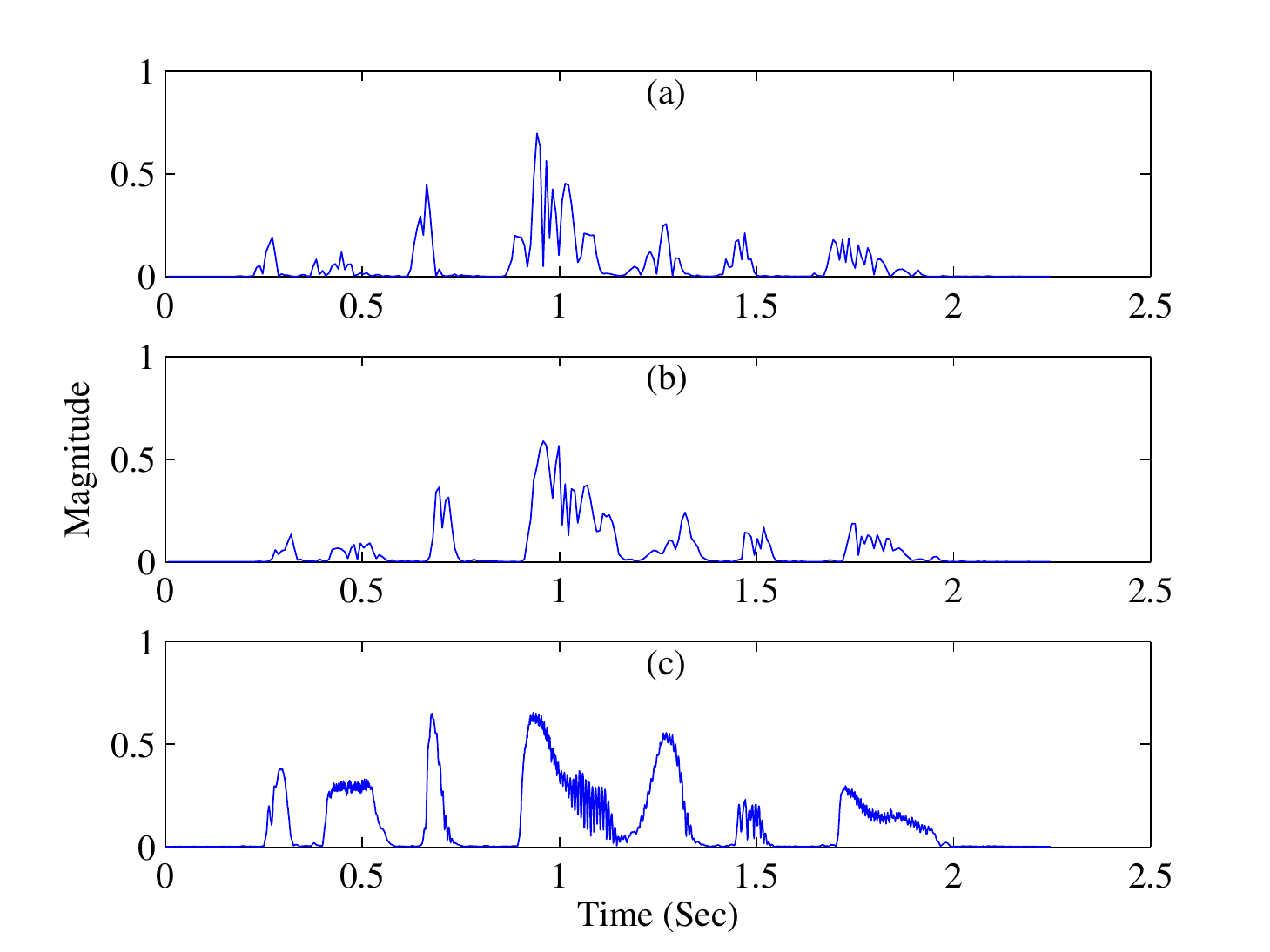}}
\caption{The speech magnitudes of (a) DWT coefficients, (b) DTCWPT coefficients, and (c) 2-stage DTCWPT coefficients in the wavelet subband. The peaks indicate the speech existence, and the magnitudes are positively correlated with the probability to be detected.}
\end{figure}

\subsection{Optimization of key parameters in speech model}

\medskip
In this study, a one-sided generalized Gamma distribution\citep{erkelens2007minimum} can be applied to describe the speech magnitude priori in DTCWPT domain 
\begin{equation}
p_{A}(a)  = \frac{\gamma\beta^{\mu}}{\Gamma(\mu)}a^{\gamma\mu-1}exp(-\beta a^{\gamma}),\beta , \gamma , \mu >0,a \geq 0
\end{equation}  
where $\Gamma(.)$ is the gamma function, $\gamma$ is usually chosen as 1 or 2, leading to bimodal or super-Gaussian priors. $\beta$ and $\mu$ are two shape parameters, and $a$ is the speech magnitude. The relations between $\gamma$, $\mu$, and $\beta$ can be described by \citep{erkelens2007minimum}
\begin{equation}
\sigma_{A}^{2} = \left(\frac{1}{\beta}\right)^{2/\gamma}\frac{\Gamma(\mu+\frac{2}{\gamma})}{\Gamma(\mu)}; 
\bar{A} = \left(\frac{1}{\beta}\right)^{1/\gamma}\frac{\Gamma(\mu+\frac{1}{\gamma})}{\Gamma(\mu)}
\end{equation} 
where $\sigma_{A}^{2} = E[a^{2}]$, and $\bar{A} = E[a]$. In (6) when $\gamma=2$, the component \(\Gamma(\mu+1/\gamma)/\Gamma(\mu)\) cannot be expressed analytically, and accordingly there is no closed-form solution. Therefore, a high-order statistic (i.e., kurtosis) $Kur$ is applied to estimate the parameter $\mu$. With some algebra, a concise kurtosis of the generalized gamma distribution can be given as  
\begin{equation}
Kur = \frac{(\mu+2)(\mu+3)}{\mu(\mu+1)},  \; \gamma =1
   \nonumber  
\end{equation}
\begin{equation}
     = \frac{\mu+1}{\mu} ,   		\;		\gamma =2
\end{equation} 
where $Kur = E[\lbrace a-\bar{A}^{2}\rbrace^4]/\lbrace E[\lbrace a-\bar{A}^{2}\rbrace^{2}]\rbrace^{2}$. In (6) it reveals that the three parameters cannot be estimated based on two equations. Therefore, it is necessary to fix the value of \(\gamma\).

To obtain an optimized speech model, 60 speech utterances from NOZEUS \citep{hu2007subjective}, and CMU speech databases \citep{langner2004creating} have been used to pre-learn the parameters of the generalized gamma distribution in the 2-stage DTCWPT domain. Speech examples range from 2 seconds to 5 seconds, and includes two males and two females. The Kullback-Leibler divergence for each band, defined as $D_{KL}(p_{A}\|h_{A})=\sum_{a}p_{A}(a)log \frac{p_{A}(a)}{h_{A}(a)}$, is applied to assess how well each statistical model explains the distribution of the speech components. $p_{A}(a)$ is the estimated speech distribution in the DTCWPT subband based on the clean speech corpus, and $h_{A}(a)$ is the histogram of the coefficients at the corresponding WPT subband. 

By (6) and (7), the parameters $\beta$ and $\mu$ of the statistical distribution fitting of each speech utterance can be estimated when \(\gamma\) is known to be 1 or 2. Figure 3 (a) shows $D_{KL}$ of the estimated models and speech distributions at all WPT subbands. The results demonstrate that the model with $\gamma=2$ has obtained lower $D_{KL}$ values than that of $\gamma=1$. Therefore, $\gamma=2$ is selected for the generalized gamma model in this study. 

\begin{figure}[!hbt]
\centering
\subfigure{\centerline{\includegraphics[scale=0.8]{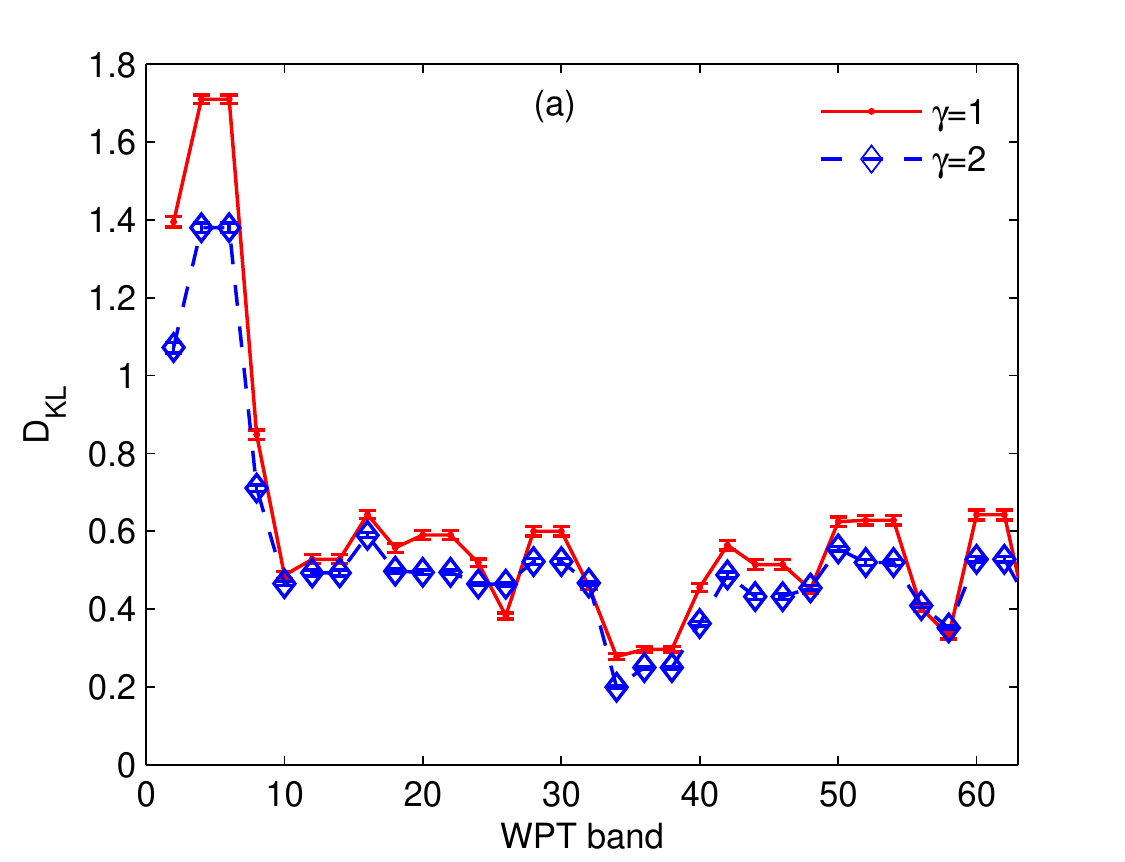}}}
\subfigure{\centerline{\includegraphics[scale=0.8]{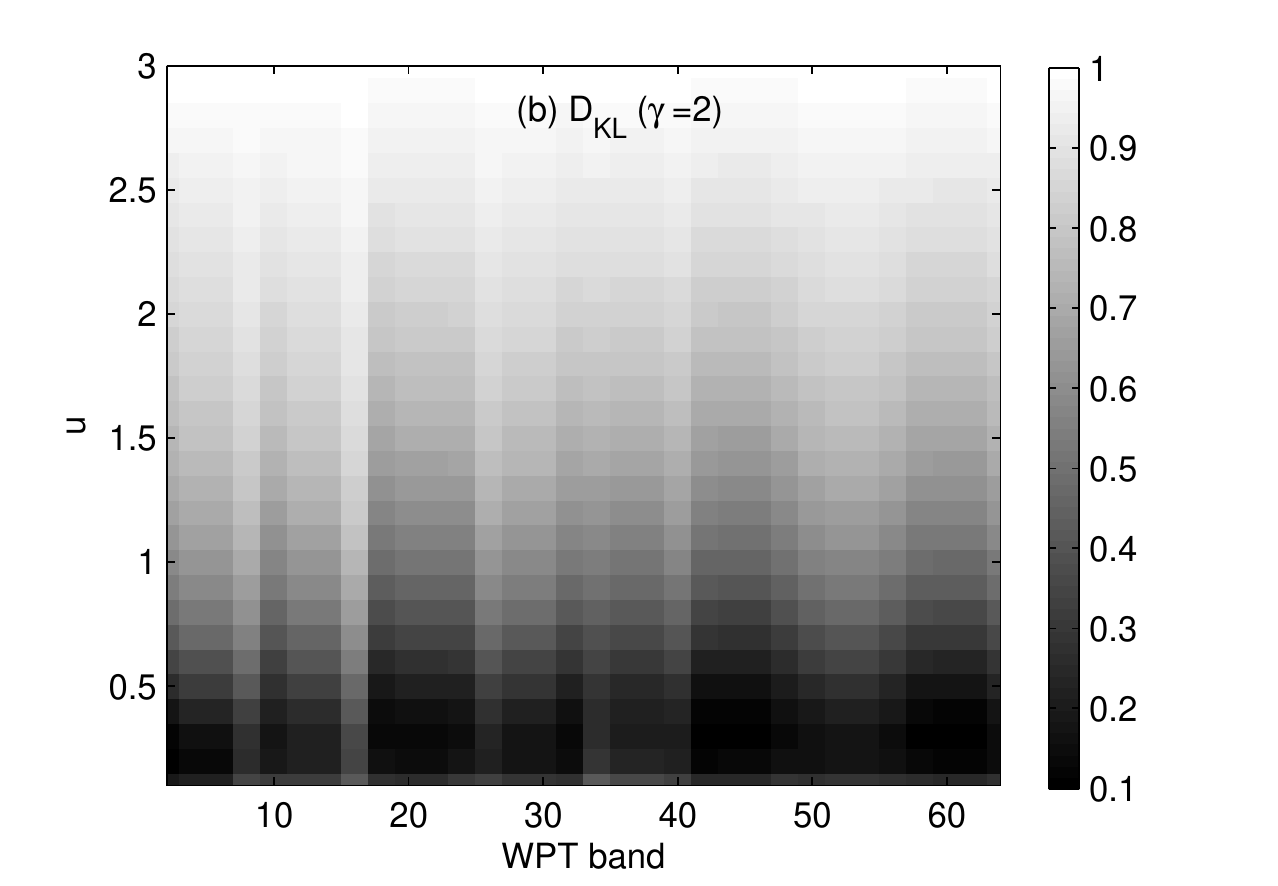}}}
\caption{(a)the divergence values $D_{KL}$ of the statistical speech model and the speech component distribution at 64 WPT subbands when $\gamma = 1$ and $\gamma = 2$, respectively. The statistical speech model is obtained by estimating $\mu$ and $\beta$ from speech samples through (6)and (7); (b)the divergence values $D_{KL}$ of the statistical speech model and the speech component distribution at 64 WPT subbands when $\gamma = 2$. The statistical speech model is obtained by various selected $\mu$ and estimating $\beta$ from speech samples through (6).}
\end{figure}

The parameter \(\mu\) is also investigated in terms of reducing the fitting divergence \(D_{KL}\). Here we incrementally selected $\mu$ to be in the range [0, 3] with a 0.1 step value, allowing $\beta$ to be estimated by (6). Figure 3(b) shows that the values of $D_{KL}$ are much smaller when $\mu$ is in the range (0,0.5] for each WPT subband. Further optimization of $\mu$ will be discussed in the next section. 

\subsection{Improved speech presence probability estimator}

\medskip
Many approaches have been developed for estimating SPP on the basis of the $a$ $posteriori$ SNR and the spectrum magnitude \citep{gerkmann2011noise, lun2012wavelet}. As illustrated in \citep{gerkmann2008improved}, a smoothing strategy by averaging the $a$ $posteriori$ SNRs of adjacent STFT coefficients has been proven effective in enhancing SPP estimation. As mentioned previously, our proposed two-stage DTCWPT multiband filters also provide a spectrum averaging effect. An improved SPP estimator is introduced as 
\begin{equation}
\mathcal{P} = P\left\lbrace \mathcal{H}_{1} \mid X \right\rbrace 
\nonumber
\end{equation}
\begin{equation}
 = \frac{P(X \mid \mathcal{H}_{1})P(\mathcal{H}_{1})}{P(X \mid \mathcal{H}_{1})P(\mathcal{H}_{1})+P(X \mid \mathcal{H}_{0})P(\mathcal{H}_{0})} = \frac{\Lambda}{1+\Lambda}
\end{equation}
where $X$ refers to the magnitude of DTCWPT coefficients in the $l$th subband, and $\mathcal{H}_{1}$ and $\mathcal{H}_{0}$ respectively represent the presence and absence of speech. Correspondingly, the generalized likelihood ratio (GLR) $\Lambda$ is defined as 
\begin{equation}
\Lambda = \frac{\kappa}{1-\kappa}\frac{P(X \mid \mathcal{H}_{1})}{P(X \mid \mathcal{H}_{0})}
\end{equation}
where $\kappa = P(\mathcal{H}_{1})$ is the $a$ $priori$ SPP, assumed to be 0.5. In the absence of speech $\mathcal{H}_{0}$, by assuming a complex Gaussian noise model, $P(X \mid \mathcal{H}_{0})$ turns out to be
\begin{equation}
p(X\mid \mathcal{H}_{0}) = \frac{1}{\sigma_{W}^{2}}exp\left( -\frac{\mid X\mid ^{2}}{\sigma_{W}^{2}}\right) 
\end{equation}  
where $\sigma_{W}^{2} = E{\mid X\mid ^{2}}$. In the presence of speech $\mathcal{H}_{1}$, the noisy signal is a sum of clean speech and noise. Assuming the speech, $A$, and the additive noise, $W$, being statistically independent random processes, $P(X\mid \mathcal{H}_{1})$ is found to be 
\begin{equation}
p(X\mid \mathcal{H}_{1})=p(A)\ast p(W)
\end{equation}
where $\ast$ denotes the convolution operator. Note that $p(W)=p(X\mid \mathcal{H}_{0})$, and $p(A)$ is as described in (5) when $\gamma=2$. Therefore, substituting (5) and (10) into (11), and applying again a variable substitution into (9), the likelihood ratio $\Lambda$ can be obtained by Eq.$3.462.1$ \cite{jeffrey2007table}
\begin{equation}
\Lambda = \frac{\kappa}{1-\kappa}\frac{2\beta^{\mu}}{\Gamma(\mu)}\int_{0}^{\infty}A^{2\mu-1}exp\left(-\left(\beta+\frac{1}{\sigma_{W}^{2}}\right)A^{2}+\frac{2X}{\sigma_{W}^{2}}A\right)dA 
\nonumber
\end{equation}
\begin{equation}
 = \frac{\kappa}{1-\kappa}\frac{2\beta^{\mu}}{\Gamma(\mu)}(2\beta^{'})^{-\mu^{'}/2}\Gamma(\mu^{'})exp(\frac{\gamma_{1}^{2}}{8\beta^{'}})D_{-\mu^{'}}(\frac{\gamma_{1}}{\sqrt{2\beta^{'}}})
\end{equation}
where $\beta^{'}=\beta+\frac{1}{\sigma_{W}^{2}}$, $\mu^{'}=2\mu$, and $\gamma_{1}=-\frac{2X}{\sigma_{W}^{2}}$. $D_{\mu}(\cdot)$ is a parabolic cylinder function of order $\mu$. According to (6) when \(\gamma = 2\), \(\Gamma(\mu+2/\gamma)/\Gamma(\mu) = \mu\), and hence $\beta=\mu/\sigma_{A}^{2}$. Alternatively, a concise form can be obtained 
\begin{equation}
 \Lambda = \frac{4\kappa}{1-\kappa}\mu^{\mu}\frac{\Gamma(2\mu)}{\Gamma(\mu)}(\mu+\xi)^{-\mu}exp(\frac{\zeta}{2(1+\mu/\xi)})D_{-2\mu}(\sqrt{\frac{2\zeta}{1+\mu/\xi}})
\end{equation}
where \(\zeta = X^{2}/\sigma_{W}^{2}\) is the $a$ $posteriori$ SNR, and \(\xi = \sigma_{A}^{2}/\sigma_{W}^{2}\) is the \(a\) \(priori\) SNR. This GLR incorporates the speech model parameter \(\mu\), as well as the general parameters $\zeta$ and $\xi$ which are often used in the speech magnitude estimators.

To investigate the impact of \(\mu\) on SPP, we fix the value of $\xi$ at 40dB \cite{gerkmann2008improved}, and let the value of \(\zeta\) sweep from -15 dB to 15 dB. The calculated values of SPP at different values of \(\mu\) are shown in Fig.4.

\begin{figure}[!hbt]
\vspace{-2mm}
\centerline{\includegraphics[scale=0.75]{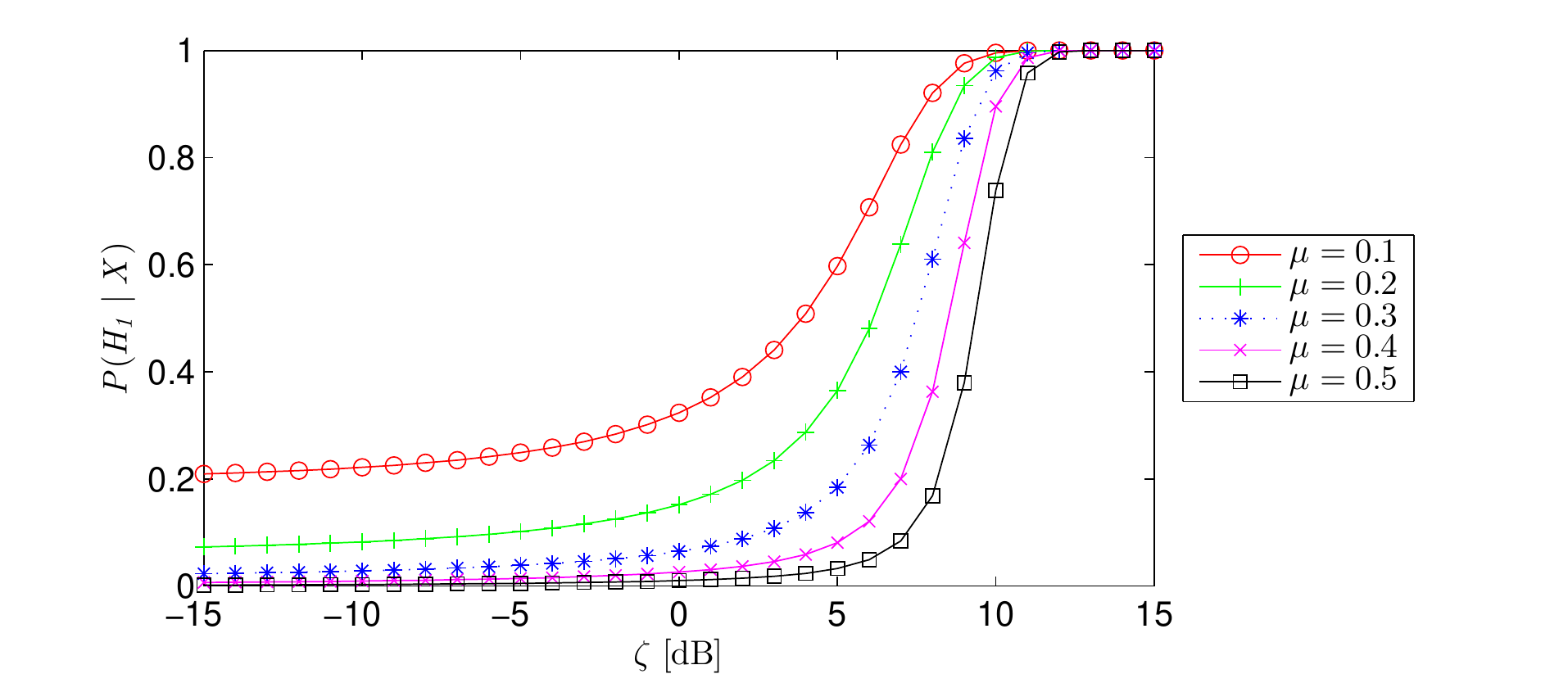}}
\caption{SPP at different a $posteriori$ SNRs $\zeta$ according to various $\mu$ with a fixed $priori$ SNR $\xi = 40dB$, according to (8) and (13).}
\end{figure}

The results indicate that at low SNRs, better performance of the SPP estimator is achieved when $\mu$ increases. For instance, when the $a$ $posteriori$ SNR is -10 dB, the SPP value with $\mu =0.5$ almost approaches zero. In other words, it detects speech with zero probability when SNR $=-10 dB$. In contrast, $P(\mathcal{H}_{1}\mid X) = 0.2$ when $\mu=0.1$, which consequently introduces noise residuals. It also reveals that when \(\mu\) is larger than 0.3, all the proposed SPP estimators have acceptable performances at low SNRs. However, when the SNR is within the range [0dB 10dB], the SPP estimator with smaller \(\mu\) has a higher probability of picking out speech components. Considering the conclusion derived in the last section that $\mu$ should be in the range (0 0.5], $\mu$ is selected as 0.3 in this study. 

With optimized \(\gamma\) and \(\mu\), the improved SPP estimator in the DTCWPT domain has been evaluated using a noisy sample, in which clean speech is corrupted by car noise at -5 dB SNR. Another SPP estimator \citep{gerkmann2008improved} is also introduced to compare with our proposed algorithm. The results in Fig.5 show that the developed SPP estimator accurately detects the speech components and excludes noise component interference well (i.e., the color bar refers to the probability of speech existence) in the low SNR scenario. Comparatively, Gerkmann's SPP estimator is more affected by noise components. Another advantage of the proposed SPP estimator is that it is highly selective on speech components. For example, the two circled speech components in Fig. 5(c) fail to be detected in Fig.5(d).
   
\begin{figure}[!hbt]
\vspace{-2mm}
\centerline{\includegraphics[scale=1]{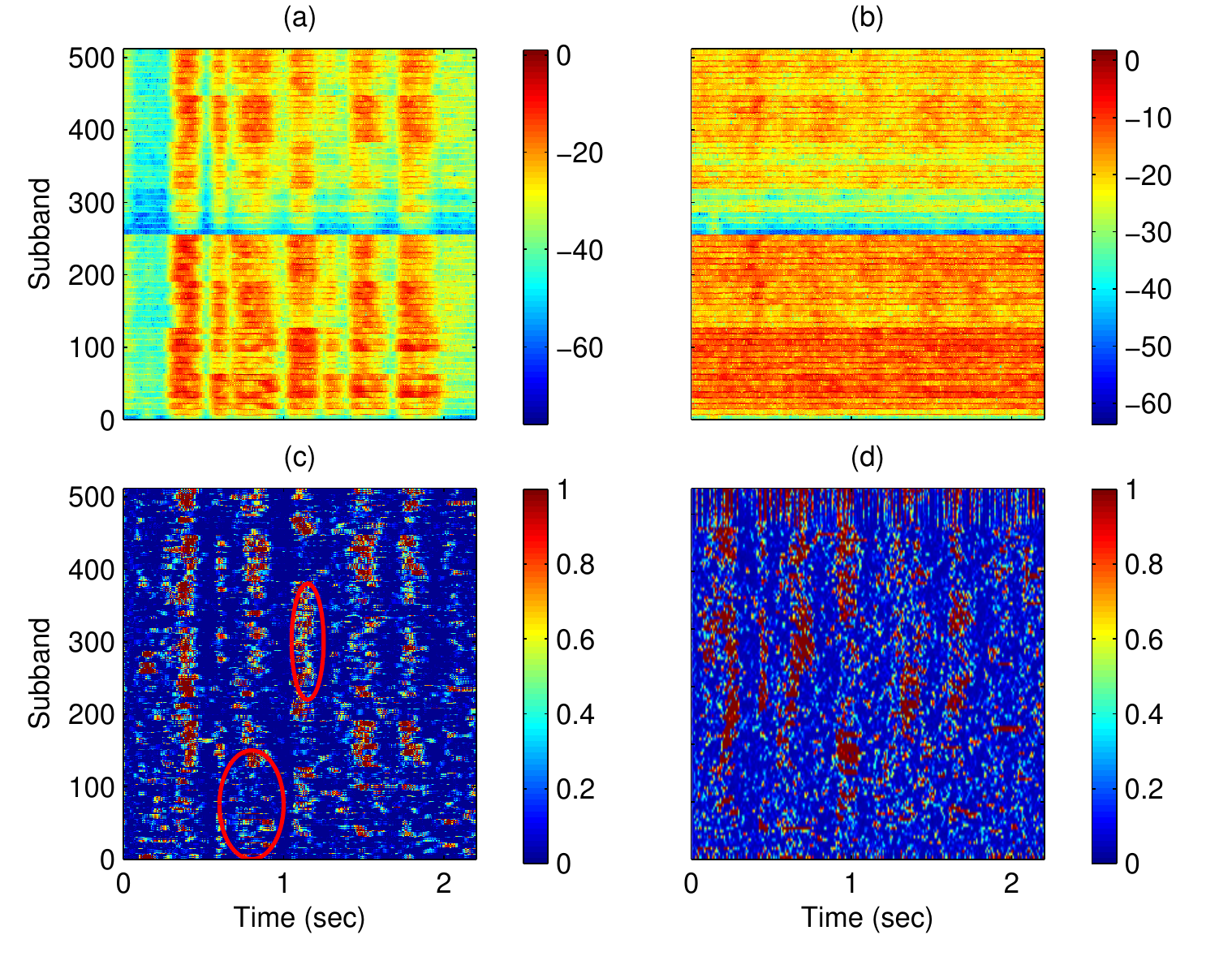}}
\caption{The spectrogram of (a) clean speech, (b) noisy speech (corrupted by car noises at -5dB SNRs) in DTCWPT domain, and SPP estimation by (c) our proposed algorithm and (d) Gerkmann's method\citep{gerkmann2008improved}.}
\end{figure}

\subsection{Statistical model for speech magnitude estimator}

\medskip

A more comprehensive SPP based speech estimator \cite{cohen2001speech} can be described by 
\begin{equation}
\widehat{A}  = \mathcal{P} \cdot E\left\lbrace A\mid X,\mathcal{H}_{1}\right\rbrace+(1-\mathcal{P})\cdot E\left\lbrace A\mid X,\mathcal{H}_{0}\right\rbrace 
\end{equation} 
where $\mathcal{P}$ is the SPP. $E\left\lbrace A\mid X,\mathcal{H}_{1}\right\rbrace$ is an MMSE estimator. An equivalent expression can be described as:
\begin{equation}
E\lbrace A\mid X,\mathcal{H}_{1}\rbrace  = \frac{\int_{0}^{\infty}ap_{X \mid A}\left(X\mid a \right)p_{A}\left(a\right)da}{\int_{0}^{\infty}p_{X \mid A}\left(X \mid a\right)p_{A}\left(a\right)da} = G_{\mathcal{H}_{1}}\cdot X
\end{equation}
where $G_{\mathcal{H}_{1}}$ refers to the gain function when speech is present. Consequently, the gain function for (14) is written as $\widehat{G} = \mathcal{P}\cdot G_{\mathcal{H}_{1}}$, given that $E\left\lbrace A\mid X,\mathcal{H}_{0}\right\rbrace= 0$. By substituting (5) and (10) into (15) \citep{erkelens2007minimum} when $\gamma =2$, $G_{\mathcal{H}_{1}}$ becomes    
\begin{equation}
G_{\mathcal{H}_{1}}^{(2)} = \frac{\Gamma(\nu+0.5)}{\Gamma(\nu)}\sqrt{\frac{\xi}{\zeta (\nu+\xi)}}\frac{_{1}F_{1}\left(\nu+0.5;1;\frac{\zeta\xi}{\nu+\xi}\right)}{_{1}F_{1}\left(\nu;1;\frac{\zeta\xi}{\nu+\xi}\right)}
\end{equation}   
where $_{1}F_{1}\left(a;b;x\right)$ is the confluent hypergeometric function. The parameter $\nu$ is defined as $\frac{\xi}{1+\xi}\zeta$.

In addition, a low-complexity SPP based noise power estimation method \citep{gerkmann2012unbiased} is introduced to dynamically track the noise variation locally. The decision-directed method is used to estimate the $a$ $priori$ SNRs. To estimate the initial noise variance, the first 0.15 seconds of each sample are assumed to be noise. 

A flow chart of the proposed algorithm is illustrated in Fig.6. Initially, the two-stage DTCWPT is applied to the input signal. Based on complex coefficients, the improved SPP estimator is then applied to the noisy speech signal, which can potentially improve the noise PSD estimation. The generalized MMSE estimator incorporates both the improved SPP estimation and the noise variance estimation to yield the estimated amplitudes. After two-stage inverse DTCWPT corresponding to upsampling and non-upsampling, a clean speech estimation is obtained. 

\begin{figure}[!hbt]
\vspace{-2mm}
\centerline{\includegraphics[scale=0.85]{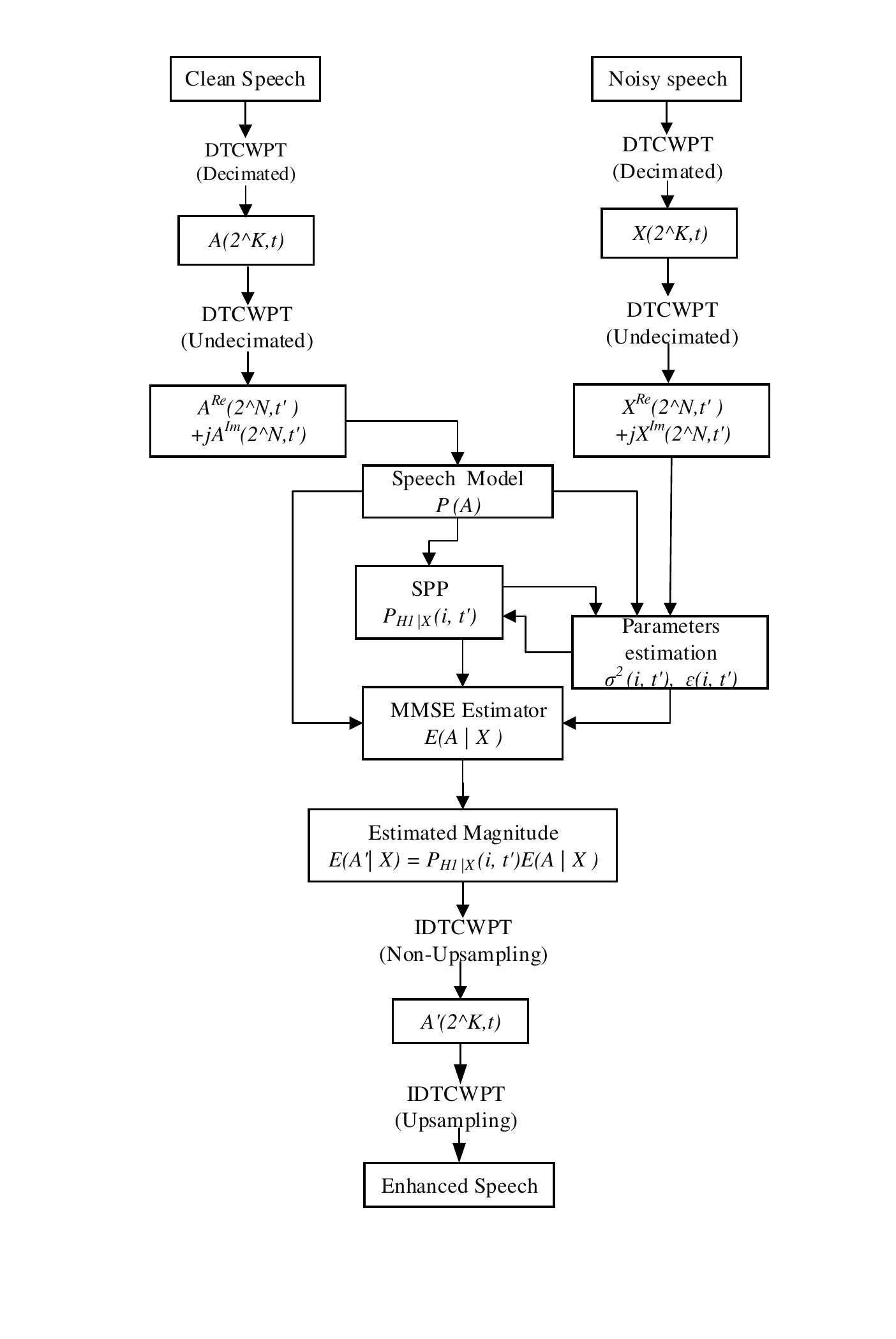}}
\vspace{-15mm}
\caption{Flow chart of the proposed two-stage DTCWPT-SPP estimator for speech enhancement. The right branch is used for the first-time development of speech model \(P(A)\), and the optimized speech model is fed into the concatenated two-stage algorithm. In the right branch, SPP estimation and parameters estimation are in a frame-based updating loop, and both of them will be applied to obtain a comprehensive magnitude estimator.}
\end{figure}

\section{RESULTS AND EVALUATION}
\medskip

The proposed algorithm has been employed in a speech enhancement framework. The noisy speech signals were synthesized by adding different noise samples, including white Gaussian noise (WGN), pink noise, babble, airport, car and train noise at different input SNRs, to randomly selected speech samples from NOIZEUS, and CMU database\citep{hu2007subjective, langner2004creating}. The noisy speech samples had an 8 kHz sampling rate at various input SNRs from -10 dB to 10 dB. The performance of the proposed algorithm has been compared with four state-of-the-art speech enhancement algorithms, including optimally modified LSA (OM-LSA)\citep{cohen2003noise}, soft masking using a posteriori SNR uncertainty (SMPO)\citep{lu2011estimators}, a posteriori SPP based MMSE estimation (MMSE-SPP)\citep{hendriks2013dft}, and adaptive Bayesian wavelet thresholding (BWT) \citep{chang2000adaptive}.

Two objective metrics, perceptual evaluation of speech quality (PESQ) and Segmental SNRs (SegSNR) as implemented in \citep{hu2008evaluation}, have been used to quantitatively evaluate the performance of the speech enhancement algorithms in this study. Figure. 7 shows the improved PESQ and SegSNR by the five algorithms for four nonstationary noises (i.e., babble, airport, car, and train) at various input SNRs. The lines respectively depict the averaged PESQ and SegSNR values for five algorithms, and the error bars are the standard deviations for 60 noisy speech samples. At low input SNRs ($<$ 0 dB), the proposed DTCWPT-SPP algorithm achieved better performances compared with the four other algorithms in terms of both PESQ and SegSNR, with a 0.3 averagely PESQ improvement and more than 1.5 dB SegSNR improvement, when compared with OM-LSA, SMPO, and MMSE-SPP. BWT algorithm achieves a good performance at -10 dB SNR, however, shows a weaker enhancement when input SNR is higher than -5 dB. At 5 dB and 10 dB input SNRs, the PESQ and SegSNR improvements of proposed DTCWPT-SPP are comparable with that of SMPO and MMSE-SPP. 

\begin{figure}
\vspace{-2mm}
\centerline{\includegraphics[scale=0.75]{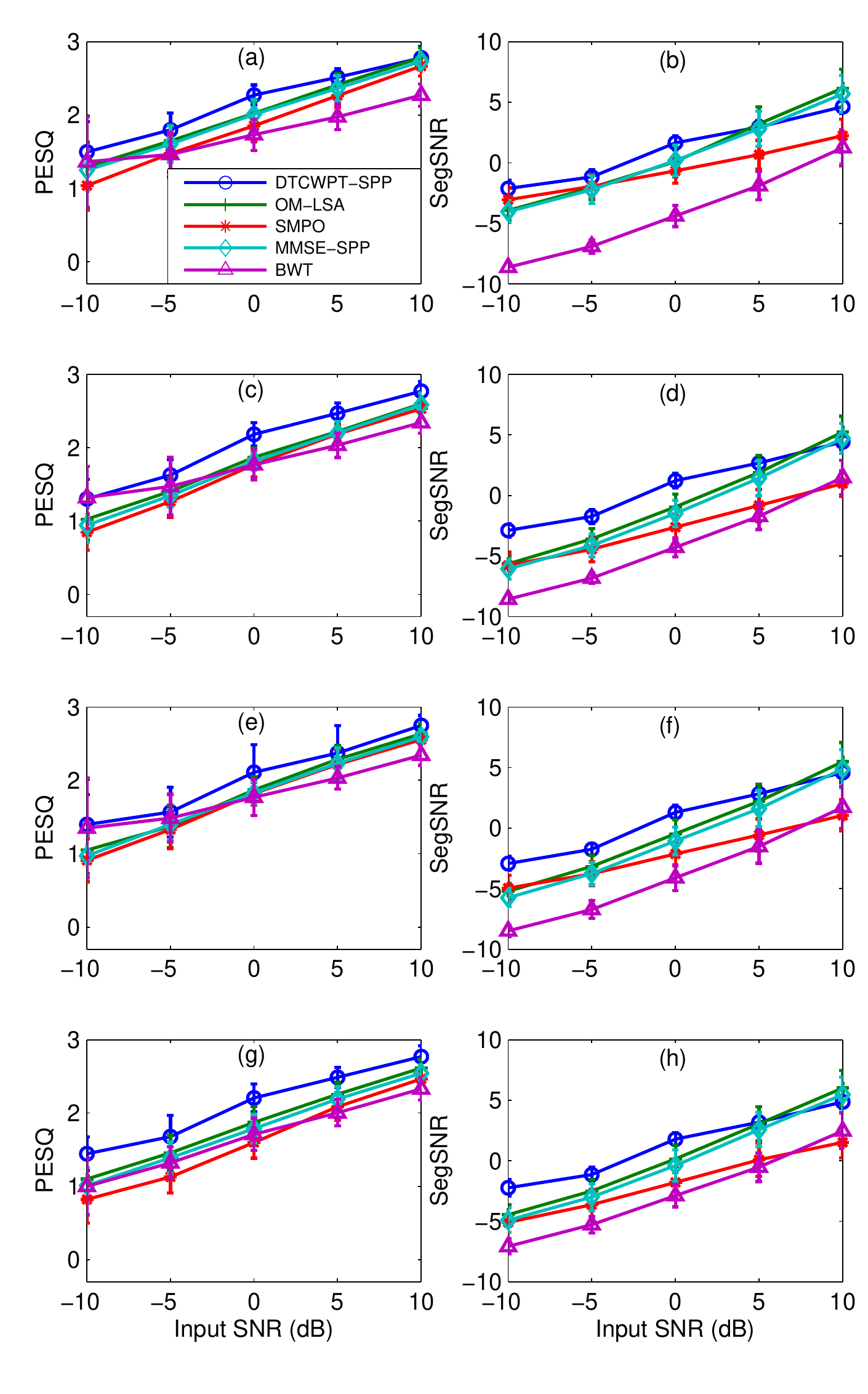}}
\caption{ PESQ and SegSNR improvements of the proposed algorithm DTCWPT-SPP and other four speech enhancement methods for four types of noise: (a)(b)babble, (c)(d)airport, (e)(f)car, and (g)(h)train at different SNR levels.}
\label{twometric}
\end{figure}

The maximum PESQ score and improved SegSNR (i.e., \(\Delta_{SNR}\)) over 60 examples at each SNR levels for four algorithms (i.e., OM-LSA, SMPO, MMSE-SPP and the proposed DTCWPT-SPP) are summarized in Table 1. Two baseline noises, referring to white Gaussian and pink noises, are also applied to corrupt the test speech corpus. Shown as the bold numbers, our proposed algorithm obtained nearly 1 dB SegSNR advantage over three other methods. It also achieved higher PESQ scores particularly with real-world noise. For example, the proposed algorithm outperformed three other algorithms by over 1 dB SegSNR improvement and 0.1 PESQ enhancement for babble noise at -10 dB SNR. The results indicate that the proposed 2-stage DTCWPT, inheriting the merit of energy-concentrated, smooth and low aliasing decomposition, demonstrates the capability to detect noise and speech components very effectively. For noisy speech corrupted by nonstationary noises at low SNRs, the improved SPP estimator more accurately detected the existence of speech components in the DTCWPT domain. With the improved SPP estimator, MMSE magnitude estimation can dynamically track the local noise variation. Therefore, the proposed DTCWPT-SPP algorithm can achieve robust performance on speech enhancement in low SNR scenarios.
\begin{center}
\begin{table*}[ht]
\footnotesize
\setlength{\tabcolsep}{4pt}
\renewcommand{\arraystretch}{0.75}
\caption{ The performance evaluation based on PESQ and SegSNRs. OM-LSA, SMPO, MMSE-SPP, and our proposed DTCWPT-SPP are implemented at 60 speech examples} 
\centering 
\begin{tabular*}{\textwidth}{ c c c c c c c c c c c } 
\hline 
&\multicolumn{2}{c}{-10 dB} &\multicolumn{2}{c}{-5 dB}  &\multicolumn{2}{c}{0 dB} &\multicolumn{2}{c}{5 dB} &\multicolumn{2}{c}{10 dB}\\  
&$\footnotesize\Delta_{SNR}$ &$\footnotesize PESQ$  &$\footnotesize\Delta_{SNR}$ &$\footnotesize PESQ$ &$\footnotesize\Delta_{SNR}$ &$\footnotesize PESQ$ &$\footnotesize\Delta_{SNR}$ &$\footnotesize PESQ$ &$\footnotesize\Delta_{SNR}$ &$\footnotesize PESQ$ \\  \hline\hline
&\multicolumn{10}{c}{White} \\ \hline
OM-LSA       &12.41  &2.77  &14.09  &3.71  &9.17  &3.79  &7.39  &3.88  &7.79  &3.73   \\
SMPO       &15.90  &2.68  &14.11  &3.31  &8.56  &3.40  &6.11  &3.44  &6.17  &3.32  \\
MMSE-SPP   &13.95  &2.77  &12.36  &3.51  &9.09  &3.57  &8.31  &3.79  &6.98  &3.59 \\ 
Proposed   &\textbf{16.19}  &2.74  &\textbf{14.63}  &3.65  &\textbf{9.34}  &3.53  &8.10  &3.78  &7.10  &3.68 \\ \hline\hline

&\multicolumn{10}{c}{Pink} \\ \hline 
OM-LSA      &6.56   &1.87  &6.55  &2.13  &6.01  &2.33  &5.72  &2.56  &5.24  &2.95   \\
SMPO      &7.37  &1.84  &6.40  &2.01  &5.22  &2.36  &4.22  &2.58  &2.88  &2.95   \\
MMSE-SPP    &6.29  &1.96  &6.20  &2.17  &5.96  &2.44  &4.95  &2.66  &5.27  &2.97 \\ 
Proposed  &6.96  &\textbf{1.97}  &\textbf{6.65}  &2.14   &5.77  &2.43  &\textbf{5.74}  &\textbf{2.69}  &4.08  &\textbf{3.00} \\\hline\hline 

&\multicolumn{10}{c}{Babble} \\ \hline 
OM-LSA      &4.18  &1.30  &3.87  &1.60  &3.62  &2.06  &3.59  &2.42  &3.10  &2.86   \\
SMPO      &4.36  &1.52  &4.07  &1.58  &4.00  &2.01  &3.62  &2.38  &4.11  &2.83   \\
MMSE-SPP  &4.59  &1.40  &3.81  &1.66  &2.22  &2.08  &1.50  &2.67  &0.76  &2.87 \\ 
Proposed  &\textbf{6.00}  &\textbf{1.64}  &\textbf{5.39}  &\textbf{1.75}  &\textbf{5.25}  &\textbf{2.11}  &\textbf{4.65}  &2.44  &\textbf{4.23}  &2.87 \\ \hline\hline

&\multicolumn{10}{c}{Airport} \\ \hline 
OM-LSA      &4.97  &1.54  &4.63  &1.75  &3.34  &2.13  &2.10  &2.78  &1.10  &2.88   \\
SMPO      &5.19  &1.64  &5.04  &1.80  &4.71  &1.98  &4.12  &2.58  &3.70  &2.80   \\
MMSE-SPP  &5.66  &1.82  &5.03  &1.87  &4.45  &2.13  &3.67  &2.72  &2.96  &2.79 \\ 
Proposed  &\textbf{6.07}  &\textbf{1.83}  &\textbf{6.01}  &1.79  &\textbf{5.15}  &\textbf{2.22}  &\textbf{4.91}  &2.72  &\textbf{4.30}  &\textbf{2.96} \\ \hline\hline

&\multicolumn{10}{c}{Car} \\ \hline 
OM-LSA     &7.45  &1.83  &6.44  &1.90  &4.78  &2.39  &3.48  &2.59  &1.45  &2.90   \\
SMPO     &6.32   &1.73  &6.14   &1.92   &5.43  &2.19  &5.12  &2.54  &4.12  &2.89   \\
MMSE-SPP  &6.55  &1.82  &6.54  &2.00  &5.50  &2.25  &4.96  &2.64  &4.64  &2.92 \\ 
Proposed  &\textbf{7.45}  &\textbf{1.89}  &\textbf{6.86}  &\textbf{2.01}  &\textbf{6.03}  &2.27  &\textbf{5.52}  &\textbf{2.71}  &\textbf{4.90}  &2.88 \\ \hline\hline

&\multicolumn{10}{c}{Train} \\ \hline 
OM-LSA     &5.65  &1.48  &5.40  &1.73  &4.85  &2.15  &4.50  &2.51  &4.06  &2.75   \\
SMPO      &5.64  &1.54  &4.91  &1.86  &3.84  &2.17  &2.43  &2.22  &1.07  &2.67   \\
MMSE-SPP  &5.88  &1.63  &5.28  &1.78  &4.77  &2.07  &4.28  &2.48  &3.71  &2.76 \\ 
Proposed  &\textbf{6.26}  &\textbf{1.70}  &\textbf{5.93}  &1.84  &\textbf{4.90}  &\textbf{2.26}  &4.06  &\textbf{2.53}  &3.64  &\textbf{2.89} \\ 
\hline 
\end{tabular*}
\label{tab:avrimprove} 
\end{table*}
\end{center}
A subjective observation in Fig. 8 demonstrates the spectrogram of clean speech, noisy speech (with car noise at -5 dB SNRs) and the enhanced speech generated by applying the four algorithms (i.e., OM-LSA, SMPO, MMSE-SPP and the proposed DTCWPT-SPP). The proposed algorithm better preserves the speech contents and more effectively removes the background noise. In Fig.8(c), it can be seen that the speech components around 1 kHz are retained, whereas the other three algorithms fail to preserve these components as shown in Fig. 8(d)(e)(f). Along the time axis around 1.6 seconds (as circled in Fig. 8(c)), the three other algorithms overcut the noise and directly eliminates the speech components. Comparatively, the proposed algorithm successfully retains most of the frequency features in this time period. Because the improved SPP accurately detects the speech components, our algorithm reconstructs the speech segments with a clear boundary as shown in the original speech spectrogramm. By taking advantages of the generalized gamma speech model and developed SPP estimator, the proposed two-stage DTCWPT algorithm obtains well-separated speech components from noise, which helps to recovery more frequency details (as circled in Fig. 8(c)) than OM-LSA or SMPO.  

\begin{figure}[!hbt]
\vspace{-2mm}
\centerline{\includegraphics[scale=0.9]{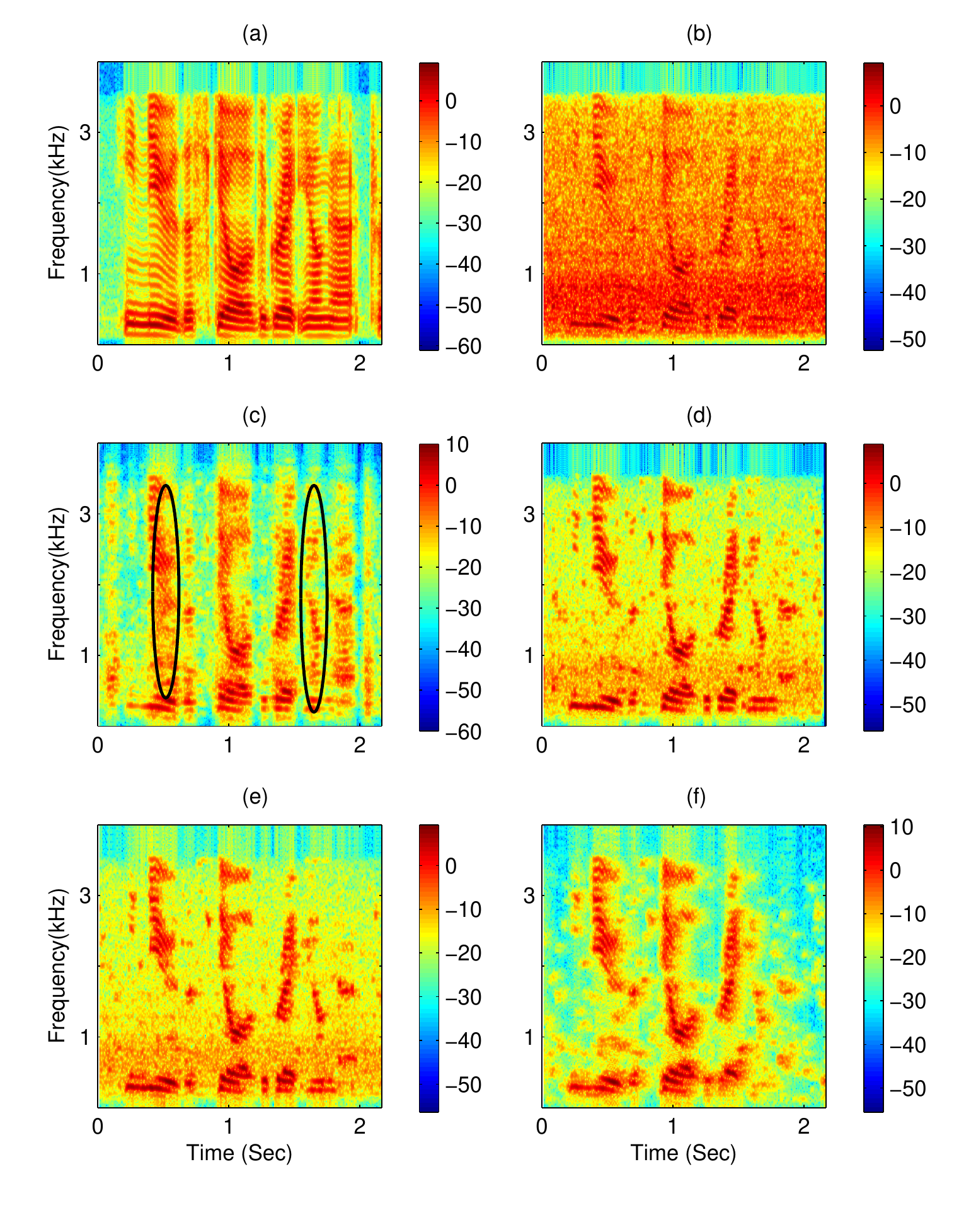}}
\caption{Spectrum of (a) clean speech , (b) speech corrupted by nonstationary car noise with -5dB SNRs,  and the enhanced speech by (c) proposed algorithm, (d) OM-LSA(Cohen2004), (e) SMPO(Louiz), and (f) MMSE-SPP(Gerkmann).}
\vspace{-5mm}
\label{subober}
\end{figure}

\section{CONCLUSION}

\medskip

In this paper, we investigated how the amplitude of subband coefficients on the basis of DTCWPT can be employed for low SNR noisy speech enhancement. A two-stage DTCWPT has been developed to obtain redundant decomposition of the noisy speech, retaining both the temporal and spectral resolutions. Based on the shift-invariant DTCWPT decomposition, a one-side gamma distribution of speech is optimized, and this optimized model is applied to derive an SPP-incorporated speech estimator. The improved SPP estimator can achieve a great performance on high noise level speech detection, and effectively eliminate noise components. The SPP incorporated MMSE estimator based on the optimized speech model was obtained. Validation results showed that our proposed algorithm achieved a considerable performance improvement on enhancing speech quality with respect to PESQ score and SegSNRs. The proposed algorithm outperformed four state-of-the-art speech enhancement algorithms at low SNRs. Further work will be done in applying the proposed DTCWPT-SPP algorithm to improve speech intelligibility for hearing protection devices used in high noise environments.

  \newpage
 
 \begin{center}

 \large{Figure Captions}
 
 \end{center}
 
 \noindent
 Figure 1. The concatenation of the (a) decimated and (b) undecimated DTCWPT. For each level decomposition, the down sampling reduces the data size into half length in decimated DTCWPT, whereas no down sampling for undecimated DTCWPT.
 
 \smallskip
 
 \noindent
 Figure 2. The speech magnitudes of (a) DWT coefficients, (b) DTCWPT coefficients, and (c) 2-stage DTCWPT coefficients in the wavelet subband. The peaks indicate the speech existence, and the magnitudes are positively correlated with the probability to be detected.
 
  \smallskip

 \noindent
 Figure 3. (a)the divergence values $D_{KL}$ of the statistical speech model and the speech component distribution at 64 WPT subbands when $\gamma = 1$ and $\gamma = 2$, respectively. The statistical speech model is obtained by estimating $\mu$ and $\beta$ from speech samples through (6)and (7); (b)the divergence values $D_{KL}$ of the statistical speech model and the speech component distribution at 64 WPT subbands when $\gamma = 2$. The statistical speech model is obtained by various selected $\mu$ and estimating $\beta$ from speech samples through (6).
 \smallskip
 
 \noindent
 Figure 4.  SPP at different a $posteriori$ SNRs $\zeta$ according to various $\mu$ with a fixed $priori$ SNR $\xi = 40dB$, according to (8) and (13). 
 \smallskip
 
 \noindent
 Figure 5.  The spectrogram of (a) clean speech, (b) noisy speech (corrupted by car noises at -5dB SNRs) in DTCWPT domain, and SPP estimation by (c) our proposed algorithm and (d) Gerkmann's method\citep{gerkmann2008improved}.
 
 \smallskip
 
 \noindent
 Figure 6. Flow chart of the proposed two-stage DTCWPT-SPP estimator for speech enhancement. The right branch is used for the first-time development of speech model \(P(A)\), and the optimized speech model is fed into the concatenated two-stage algorithm. In the right branch, SPP estimation and parameters estimation are in a frame-based updating loop, and both of them will be applied to obtain a comprehensive magnitude estimator.  
 \smallskip
 
 \noindent
 Figure 7.  PESQ and SegSNR improvements of the proposed algorithm DTCWPT-SPP and other four speech enhancement methods for four types of noise: (a)(b)babble, (c)(d)airport, (e)(f)car, and (g)(h)train at different SNR levels.
 \smallskip
 
 \noindent
 Figure 8.  Spectrum of (a) clean speech , (b) speech corrupted by nonstationary car noise with -5dB SNRs,  and the enhanced speech by (c) proposed algorithm, (d) OM-LSA(Cohen2004), (e) SMPO(Louiz), and (f) MMSE-SPP(Gerkmann).
 \smallskip

 \end{document}